# Selective Trapping of Hexagonally Warped Topological Surface States in a Triangular Quantum Corral

Mu Chen[1,2], Yeping Jiang[3]*, Junping Peng[1], Huimin Zhang[1], Cui-Zu Chang[1,7], Xiao Feng[1], Zhenguo Fu[4], Fawei Zheng[4], Ping Zhang[4], Lili Wang[1,5], Ke He[1,5], Xu-Cun Ma[1,5]*, and Qi-Kun Xue[1,5,6]*

[1] *Department of Physics, Tsinghua University, Beijing 100084, China*

[2] *Beijing Institute of Aeronautical Materials, Beijing 100095, China*

[3] *Key Laboratory of Polar Materials and Devices (MOE), Department of Optoelectronics, East China Normal University, Shanghai 200241, China*

[4] *Institute of Applied Physics and Computational Mathematics, Beijing 100088, China*

[5] *Collaborative Innovation Center of Quantum Matter, Beijing 100084, China*

[6] *Beijing Academy of Quantum Information Sciences, Beijing 100193, China*

[7] *Department of Physics, The Pennsylvania State University, University Park, Pennsylvania 16802, USA*

*To whom correspondence should be addressed to Y.P.J. (ypjiang@clpm.ecnu.edu.cn), X.C.M. (xucunma@mail.tsinhua.edu.cn) or Q.K.X. (qkxue@mail.tsinghua.edu.cn).

**The surface of a three-dimensional topological insulator (TI) hosts two-dimensional massless Dirac fermions (DFs), the gapless and spin-helical nature of which yields many exotic phenomena, such as the immunity of topological surface states (TSS) to back-scattering. This leads to their high transmission through surface defects or potential barriers. Quantum corrals, previously elaborated on metal surfaces, can act as nanometer-sized electronic resonators to trap Schrödinger electrons by quantum confinement. It is thus intriguing, concerning their peculiar nature, to put the Dirac electrons of TSS to the test in similar circumstances. Here, we report the behaviors of TSS in a triangular quantum corral (TQC) fabricated by epitaxially growing Bi bilayer nanostructures on the surfaces of $Bi_2Te_3$ films. Unlike a circular corral, the TQC is supposed to be totally transparent for DFs. By mapping the electronic structure of TSS**

**inside TQCs through a low-temperature scanning tunneling microscope in the real space, both the trapping and de-trapping behaviors of the TSS electrons are observed. The selection rules are found to be governed by the geometry and spin texture of the constant energy contour of TSS upon the strong hexagonal warping in $Bi_2Te_3$. Careful analysis of the quantum interference patterns of quasi-bound states yields the corresponding wave vectors of trapped TSS, through which two trapping mechanisms favoring momenta in different directions are uncovered. Our work indicates the extended nature of TSS and elucidates the selection rules of the trapping of TSS in the presence of a complicated surface state structure, giving insights into the effective engineering of DFs in TIs.**

# INTRODUCTION

The existence of topological surface states (TSS), with their low energy behavior described by Dirac equation rather than Schrödinger equation, is ensured by the twisted bulk band topology resulting from the strong spin-orbital coupling of the material (*1, 2*). Unlike the massive electron systems having a forbidden gap, such as any trivial band insulator and even massive Dirac systems, the gapless Dirac fermion (DF) nature of TSS makes the potential barriers leaky(*3*). Furthermore, as long as time-reversal symmetry is respected, the backscattering of TSS is prohibited because of its helical nature(*3-5*), leading to the full transmission of normal incident electrons through potential barriers. This situation is more prominent for a family of three-dimentional (3D) topological insulators (TIs) ($Bi_2Se_3$, $Bi_2Te_3$, and $Sb_2Te_3$)(*6-9*), which possess a single Dirac cone in the surface state structure and only one Fermi surface on the constant energy contour (CEC) within the bulk gap, further eliminating possible backscattering channels. Thus, the two fundamental properties of TSS, gaplessness and helical spin texture, may render any conventional quantum corral(*10*) highly leaky and complete trapping of TSS impossible, similar to the case of DFs in graphene(*11*).

For a 1D barrier on the TI surface, although reflection for normal incident electrons is forbidden, the reflection at oblique angles is possible, and even dominates the tunneling process at large incident angles, approaching one at grazing angles(*12*), which is governed by the spin texture of the CEC(*13, 14*). This makes the quasi-bounding of TI surface states possible via whispering-gallery modes as observed in a circular graphene pn junction(*15-17*), or in the case of antimony by constructing a Fabry–Pérot resonator(*3, 12*), the 'infinity' of which along the 1D barriers helps to trap the TSS. However, antimony has a more complicated surface state structure(*18*), which facilitates backscattering and makes the trapping behavior of surface states more conventional. The above-mentioned approaches both utilize the high reflection probability at near-grazing angles of DFs. These intrigue the interest to construct different resonators, in which electrons can no longer keep bouncing at grazing angles, on TIs but with simple CECs.

Here we successfully constructed such resonators as triangular quantum corrals (TQCs) on the surface of a 3D TI, $Bi_2Te_3$, which has a single Dirac cone and only one Fermi surface on the CEC in the surface state structure within the gap(*6*). Although the trapping of DFs by TQCs seems impossible because of the above argument, our experiments show well-defined interference patterns in some energy ranges of the TSS band. The trapping of Dirac electrons is found to be assisted by the hexagonal warping effect in the TSS, which in $Bi_2Te_3$(*19*) is much stronger than that in $Bi_2Se_3$(*20*). As shown below, the strong warping effect of TSS in $Bi_2Te_3$ yields energy-dependent trapping behaviors and selection rules of TSS in TQCs.

## RESULTS

### MBE fabrication of TQCs and the observation of trapped TSS

Our TQCs are constructed by hetero-epitaxially growing a Bi bilayer (BL) on the surface of the $Bi_2Te_3$ thin film (see Methods). The previous work(*21*) shows that at appropriate Bi coverages (typically 0.9 BL) $Bi_2Te_3$ vacancy islands surrounded by the Bi BL film can be formed. Taking an island with the side length $L$ of ~35 nm, a close scrutiny uncovers quasi-equilateral triangular vacancy islands (Fig. 1A), bounded by $\langle\bar{1}01\rangle$, $\langle 1\bar{1}0\rangle$, and $\langle 01\bar{1}\rangle$-oriented steps of the Bi BL(*21*). The depth of the vacancy island height is ~ 4.8 Å and bias-voltage independent. The defect-free atomic resolution image inside the TQC demonstrates that the Te-terminated $Bi_2Te_3$ (111) surface possesses a lattice spacing of ~ 4.4 Å, consistent with the bulk lattice constant(*22*). The orientations of the triangular sides of the vacancy island align with the $\bar{\Gamma}-\bar{K}$ directions of the $Bi_2Te_3$ surface Brillouin zone.

The Dirac energy of the pristine $Bi_2Te_3$ film locates at ~ -170 meV (supplementary Fig. S1) and is buried below the bulk valence band. Compared with pristine films, the scanning tunneling spectroscopy (d$I$/d$V$) taken at the TQC center shows a downward shift of ~ 130 meV (the blue curve in Fig. 1B), indicating the electron-doping effect of the Bi BL on the $Bi_2Te_3$ film(*21*). The Dirac energy of TSS in the uncovered region should be around -300 meV. The presence of a Bi BL changes the boundary condition of the Bi$_2$Te$_3$ film and the Dirac energy is lifted to ~ -210 meV. Therefore, there is a mismatch between the Dirac energies across the interface along $\overline{\Gamma}-\overline{K}$ directions between Bi$_2$Te$_3$ and Bi/Bi$_2$Te$_3$ regions, in addition to the potential barriers caused by the steps. TQCs are thus formed on the surface of Bi$_2$Te$_3$ films for further study of trapping behaviors of TSS as discussed below.

We note that the peaks shown in the spectrum taken at the TQC center (Fig. 1B) indeed reveal the occurrence of quantized states, which is against our previous analysis that a TQC cannot trap massless DFs. The spatial patterns of these states were imaged by taking d$I$/d$V$ maps in the energy range from -400 meV to 500 meV with a 25-meV interval (see supplementary Fig. S2). Figure 2 shows the d$I$/d$V$ maps at some representative energies that capture the main feature of trapping behaviors of TSS. The regular patterns can be clearly seen below -250 meV or above -175 meV, and no pattern can be seen in between (except a bulk-state related pattern at -240 meV, which can be seen in Fig. 1B and more clearly in Fig. 4A). The patterns in the energy window from -175 meV to -100 meV are relatively weak and get much stronger above -75 meV.

**Modeling of the interference patterns**

A simple particle-in-a-box model(*23*) is employed and diagrammed in Figs. 3A and 3B to simulate the above patterns and to elucidate the exact nature of these trapped states as well as the TSS selection rules in such a TQC. Provided the reflections are appreciable, an electron in a TQC with a side length of $L$ can be possibly trapped by successive reflections at the triangular barrier walls, generating six wave vectors $k_1$, $k_2$, $k_3$, $k_4$, $k_5$, $k_6$, which have the $c_{3v}$ symmetry and are related to each other by in-plane 120° rotations and by reflections with respect to $\overline{\Gamma}-\overline{K}$ (Figs. 3B and 3C). Taking the coordinates as shown in Fig. 3A and by imposing the boundary condition, these superpositions of momentum states are then quantized, such as $k_1 = \left(\frac{2\pi}{3L}(2n-m), \frac{2\pi}{\sqrt{3}L}m\right)$ with a magnitude of $k_{nm} = \frac{4\pi}{3L}\sqrt{n^2+m^2-nm}$. Therefore, by matching the interference patterns at specific energies, a series of ($E_{nm}$, $k_{nm}$) states can be obtained to form the ($n$, $m$) quasi-bound states of TSS in the TQC. Because of the degeneracy caused by symmetry, only the states of $n \geq 2m$ are considered, where $n$, $m$ are positive integers. Note that the defect-free condition is indispensable in this work. Otherwise, the existence of point scatterers will invalidate the above model.

The simulated eigenmodes are shown in the inserts of Fig. 2, which fit the interference patterns best at different energies. The patterns at and below -250 meV are found (see the discussion section) to be caused by the band-bending of the bulk states, except those at -275 and -300 meV, which are attributed to the (2, 1) and (3, 1) states, respectively. The patterns from -175 to -100 meV all have higher $m$-indexes ($m \geq 2$) but are relatively weak, much weaker than the spectral intensity of the Bi BL, especially the one at -175 meV which is nearly indiscernible. At energies above -75 meV, the interference patterns get stronger and switch to low $m$-indexes ($m = 1$).

In the diagram of Fig. 3C, the vector series corresponding to low- or high $m$-indexed states are indicated on the energy-dependent CECs of TSS in $Bi_2Te_3$, as well as the complicated spin texture due to the strong hexagonal warping. For the ($n$, 1) states, the vectors $k_{vi}$ are close to valley positions in $\overline{\Gamma} - \overline{K}$ directions on the warped CECs (the wave vectors for $m = 0$ states are exactly along $\overline{\Gamma} - \overline{K}$ directions but the corresponding wave functions are zero inside the TQC). For states with higher $m$-indexes, the vectors $k_{ti}$ are close to tip positions in $\overline{\Gamma} - \overline{M}$ directions on the CECs. Note that vectors of the ($n = 2m$, $m$) states are exactly along $\overline{\Gamma} - \overline{M}$ directions. Thus the scattering process of the surface states inside the TQC reveals a strong energy dependent trapping of TSS, selecting momenta in different directions on the CECs, which, due to the strong warping effect, have quite different magnitudes and spin orientations.(*19*)

## DISSCUSSION

### The identification of trapped states below the highest valence subband

Figure 4A provides the spatially resolved energy levels of trapped states by taking the d$I$/d$V$ spectra along the line across the TQC (thick arrow in Fig. 3A). $E_c$ (-75 meV) denotes the conduction band minimum as identified by the d$I$/d$V$ spectra (Fig. 1B). By using the pattern in Fig. 4A, the correction of the energies of the trapped states can be made. In addition, two trapped surface states (2, 1) and (3, 1) are identified. The spots at or below -240 meV, which correspond to the peaks in the d$I$/d$V$ spectra of different sized TQCs (35 and 17 nm, see supplementary Fig. S3), reflect the bulk valence subbands. The highest valence subband (denoted as $E_v \sim$ -240 meV) also shows as a peak ($\sim$ -110 meV) in the d$I$/d$V$ spectrum on the pristine $Bi_2Te_3$ film (Fig. 1B) but with an energy shift caused by different doping. These bulk related patterns are identified to be the density variation of valence band states caused by band bending in the lateral direction near the steps, as noted above the Bi BL has a strong electron-doping effect on $Bi_2Te_3$. The (2, 1) states centered at -280 meV in Fig. 4A, corresponding to the pattern at -275 meV, obeys a sinusoidally spatial intensity distribution, which is in contrast to other patterns below $E_v$. The (3, 1) pattern at $\sim$ -300 meV (-320 meV after correction) doesnot have a node in the TQC center because of its weak intensity and the relatively stronger density variation of bulk states caused by band bending.

**Seven different energy regions in the band structure of $Bi_2Te_3$**

Thus, by a careful investigation of Figs. 4A and 4B, as well as the interference patterns at different energies (Fig. 2 and Fig. S2), seven energy regions with distinct trapping behaviors or interference patterns are clearly identified. The first region is below -320 meV, in which, except for the patterns caused by band bending, no trapped state can be observed. The second region is from -320 meV to $E_v$. Besides the patterns coming from the band-bending effect, two trapped states, (3, 1) at -320 meV and (2, 1) at -280 meV, are found, respectively. The (3, 1) state has a clover shape that cannot be explained by the band-bending effect. The pronounced peak of the (2, 1) state occurs in the d$I$/d$V$ spectra of the TQC ($L$ = 35 nm) but does not appear in that of a smaller TQC ($L$ = 17 nm, see supplementary Fig. S3) and that of the pristine film, indicating its trapped nature. The third region is from $E_v$ to $E_w$ (-175 meV), in which no pattern appears (Figs. 2 and 4B). Here $E_w$ denotes approximately the energy above which the CEC of TSS becomes concave. The fourth region is from $E_w$ to $E_c$ (-75 meV), in which there are very weak patterns (Fig. 2). These weak patterns all match with the states of ($n, m \geq 2$).

The fifth region is from $E_c$ to 75 meV. In this region, the patterns are stronger but seem mixed, such that there is no clear spots or nodes in the spatial map of energy levels (Fig. 4B), although the dominating contribution comes from the ($n$, 1) modes. The pattern mixing is possible because of the level broadening caused by the finite lifetime of the trapped states and the unequal spaced energy levels of ($n$, $m$) states (some states with different $n$ and different $m$ may be very close). This means that the trapping probabilities of different channels are competing in such an energy range. The sixth region is from 75 to 400 meV, in which strong ($n \geq 10$, 1) patterns are obvious at a glance. The indexing of trapped states can be double-checked by the symmetry of the ($n$, $m$) states. This can be done by noting that ($3i+2$, $3j+1$) and ($3i+1$, $3j+2$) states ($i$, $j$ are non-negative integers) have an antinode at the center of the TQC (the dashed line in Fig. 4A). The last region is above 400 meV, where there are no more clear patterns. Note that the absence of trapped states in regions 1 and 7 is a result of the finite energy range of TSS.

**Detailed nature of the trapped TSS in four regions (6, 5, 4, 2)**

Based on the above observation and discussion, the schematic band structure of our Bi$_2$Te$_3$ thin film is shown in Fig. 4C. Figure 4D gives a summary of the indexed states trapped by the TQC ($L = 35$ nm). The energies of the trapped states are all corrected by Fig. 4A, except those above 250 meV (half filled squares). Here we may focus on the details of the above-mentioned trapping behaviors (in regions 6, 5, 4, and 2). For the ($n \geq 6$, 1) states (regions 6 and 5, filled squares in Fig. 4D), the vectors $\mathbf{k}_{vi}$ are close to valley positions in $\overline{\Gamma}-\overline{K}$ directions on the warped CECs (Fig. 3C). Thus, the derived dispersion by fitting (red solid line) gives the Fermi velocity of Bi$_2$Te$_3$ thin films in the $\overline{\Gamma}-\overline{K}$ direction, $v_{F1} \sim 4.9 \times 10^5$ m/s, consistent with the result from step-edge scattering(*4*). At higher energies, the step-edge scattering, which may have different scattering vectors, starts to dominate the interference pattern near the edges (above 200 meV in Fig. S2), resulting in the inaccuracy of counting the nodes of the trapped modes (the deviation of half-filled squares from the linear dispersion of TSS). In addition, the energy correction by Fig. 4A is impossible for states above 250 meV. The extrapolation of the fitting yields a Dirac energy of ~ -300 meV, in excellent agreement with our d$I$/d$V$

spectra(*21*). These trapped ($n ≥ 6$, 1) states are no doubt from TSS because for the 35-nm TQC the (6, 1) state instead of the ground states (2, 1) of the conduction band parabola lies close to the conduction band minimum. It is also not likely that a Bi overlayer can trap the bulk states that extend across the whole thickness of the film.

For the higher *m*-indexed ($n$, $m ≥ 2$) states (region 4, hollow squares from -175 to -100 meV in Fig. 4D), the vectors $k_{ti}$ are close to tip positions in $\overline{\Gamma}-\overline{M}$ directions on CECs (Fig. 3C). These in-gap high-*m* states are exclusively from TSS. The slower Fermi velocity $v_{F2} \sim 3.6×10^5$ m/s is in agreement with a sudden decrease of the slope of the TSS band along $\overline{\Gamma}-\overline{M}$ directions above $E_w(6)$. Note that the trapping of these high-*m* states involves scattering between states on the CEC that are nearly time-reversal symmetric (TRS), especially for the ($n = 2m$, $m$) states. This situation is different from the interference pattern of TSS scattered by a single line-defect(*4, 24, 25*), where the nesting vector ($q_2$) connecting, for example, $k_{t2}$ and $k_{t3}$, induces a strong interference pattern. In the TQC case, the absence of scattering vector $q_3$ ($q_3$ connects two nearly TRS counterparts) will render the trapping states ($n$, $m ≥ 2$), especially the ($n = 2m$, $m$) states, of TSS impossible even in the presence of strong reflections $q_2$. There must be a TRS-breaking mechanism for the appearance of these high *m*-indexed states.

The (2, 1) and (3, 1) states below $E_v$ (region 2) are also from TSS by using the argument for the states above $E_c$ and also by noting that there is only one (2, 1) state despite the multiple valence subbands. However, the appearance of the (2, 1) state, especially its strong intensity, seems like a puzzle based on the arguments above, contradicting TRS.

## The mechanism of three kinds of trapping behaviors

Our interpretation of these three kinds of trapping (in regions 2, 4, 5 and 6) behaviors is as follows. First, the Dirac energy is buried well below $E_v$. This makes the trapping of TSS below $E_v$ (region 2) possible because there may exist two Fermi pockets with inverted spin textures similar to the antimony case(*6, 18*), especially for the (2, 1) state, as shown in Fig. S6, where the backscattering is no longer forbidden. The deviation of $k'_{21}$ from the TSS dispersion (Fig. 4D) results from the fact that the trapping involves now wave vectors from multiple Fermi surfaces. This scenario supports the situation that the Dirac point lies below the (2, 1) state, in agreement with our observation. The absence of the (2, 1) state for the smaller TQC ($L$=17 nm, Fig. S4) is reasonable because its energy may lie in the energy region where trapping is impossible (region 3).

The other two trapping behaviors (in regions 4, 5 and 6) are found to be caused by the strong hexagonal warping in $Bi_2Te_3$, which leads to heavily deformed CECs and modified spin textures of TSS band(*19*). The energy-dependent CECs evolve from a circle near the Dirac point to a hexagon and then to a snowflake(*6*). In the case of a concave CEC above a certain energy, there may exist spin-polarized local modes along a line defect extended in the $\overline{\Gamma} - \overline{K}$ direction(*26*). The existence of bound states near the step edge on $Bi_2Te_3$ has been reported before(*27*). Thus, the occurrence of trapped states in the forbidden gap (region 4) may be ascribed to the simultaneous appearance of localized states along the barriers (Figs. 2 and 4B, or S5). These localized states are even stronger in intensity than the interference patterns inside the TQC, and may act as a magnetic barrier that opens the channel of nearly normal reflections responsible for the trapping of ($n, m \geq 2$) states. Nonetheless, the patterns of ($2m, m$) states should still be absent because of the orthogonality of scattering vectors in the spinor basis. The TRS counterparts simply do not interfere(*28*). In fact, these ($n, m \geq 2$) patterns are all very weak, especially the almost indiscernible (4, 2) pattern at -175 meV, as shown in our experiment. The pattern at -125 meV (Fig. 2) has contributions from the (6, 2) and (6, 3) modes but with an intensity ratio of about 2 to 1, although these two modes are nearly degenerated ($k_{62}$ and $k_{63}$ are close). Their appearance, although very weak, may originate from the non-ideal shape of the TQCs in some sense, which may result in a non-perfect alignment between the TQC edges and the $\overline{\Gamma} - \overline{K}$ directions. The trapping of the (2, 1) state below $E_v$ involves backscattering between electronic states with the same spin direction, which is different from this scenario and explains the strong intensity of the (2, 1) pattern at -275 meV.

Furthermore, the complexity of the spin texture (diagrammed in Fig. 3C) due to the strong hexagonal warping(*19*) may account for the appearance and the growing intensity of the ($n \geq 6$, 1) states with energy. In the modified spin texture, the spin polarization near the $\overline{\Gamma}-\overline{K}$ directions are mostly affected, with the spin gradually canting towards the out-of-plane direction. Thus, the reflections between these valley points ($k_{vi}$) increase with energy(*13*), modifying drastically the reflectivity-incident angle relation compared to the case of TSS with a simple spin-momentum texture. Note that the wave vectors of the ($n \geq 6$, 1) states are not perfectly along $\overline{\Gamma}-\overline{K}$ directions. The trapping of these states is composed of these inter-valley reflections ($q_2$, $q_3$) and intra-valley ones ($q_1$, grazing angles). The larger *z*-component conformity of spins between the scattering wave vectors also makes the interference pattern stronger with energy(*13*). This explains the existence of a transit region 5 (from ($n$, $m \geq 2$) to ($n$, 1) domination) where the interference patterns are complicated. The region 5 in our case is from $E_c$ to 75 meV. Below $E_c$, the ($n$, 1) states are missing, such as the (4, 1) and (5, 1) states. Above 75 meV, only the ($n$, 1) states are visible in the d$I$/d$V$ maps. A previous experiment(*24*) also observed an anomaly at some energy above $E_c$ in the interference patterns along the step edge. We argue this anomaly by the change of phase-shift at the boundary with energy. This can be seen more obviously by investigation of the patterns (near the edges) of 50, 75 and 100 meV (Fig. S7), which indicates a phase-shift change from 0 to π above $E_c$. This implies that the dominant scattering process responsible for the trapping of TSS changes above $E_c$, in accordance with the switch of trapped modes from ($n$, $m \geq 2$) to ($n$, 1). The deviation of the phase-shift from π below 100 meV results in an underestimation of $k_{nm}$, which explains the slight deviation of states (9, 1) and (10, 1) from the linear fitting quite well (Fig. 4D). Furthermore, the trapping of TSS above $E_c$ via bulk states is unlikely, because at low temperatures (~ 4 K) the phonon-assisted inelastic scattering between TSS and the bulk states is totally suppressed. They behave more like two independent electron systems that are weakly coupled even at room temperature, with a time scale of several picoseconds(*29, 30*).

Despite all the above-mentioned reflectivity-improving mechanisms that help to trap the DFs in TIs, the life time of the trapped states is still limited. The typical life time estimated from the peak widths (from 20 to 40 meV) ranges from 20 to 40 fs, which corresponds roughly to the reflectivity of TSS at the TQC edges from ~ 0.12 to

~ 0.35 (see Fig. S8 for details). Here only the states (*n*, 1) are discussed, because the high-*m* states are too weak. The relatively low reflectivity indicates the high transmission of TSS at the TQC edges.

**Conclusions**

To conclude, despite the promised transparency of TQCs for TSS, three situations to trap the TSS are directly visualized in our work. One is the presence of double Fermi surfaces having opposite spin textures. For $Bi_2Te_3$, this region lies below the bulk valence band maximum. The other two take advantage of the strong hexagonal warping of TSS in $Bi_2Te_3$, which leads to the existence of possibly magnetic bound states and a more complicated spin texture. While the former opens the forbidden back-scattering channels that facilitate the trapping of the high *m*-indexed (*n*, *m* ≥ 2) states of TQCs, the latter favors the trapping of the (*n*, 1) states, selecting the states near the tips ($k_{ti}$) and those close to the valleys ($k_{vi}$) on the CEC, respectively. In the situation where all the afore-mentioned conditions are absent, the TQCs are completely transparent for the TSS, in agreement with the argument presented at the beginning. This region in $Bi_2Te_3$ is from $E_v$ to $E_v$ +85 meV where the CECs of TSS are supposed to be convex, constituting a truly forbidden region to trap the TSS. Note that the (3, 1) and (4, 1) states (Fig. 4D) above the Dirac point are missing. Similar experiments can be conducted on TIs without strong hexagonal warping ($Bi_2Se_3$ or $Sb_2Te_3$), which may have different trapping behaviors and may further validate our interpretation of current results. Thus, our findings elucidate the trapping and the de-trapping behaviors of TSS in the presence of complicated CEC geometry and spin textures, giving insight into the manipulation of TSS, which is especially intriguing owing to their massless, relativistic and spin 1/2 DF nature.

**MATERIALS AND METHODS**

**MBE growth and scanning tunneling microscopic (STM) characterizations**

The main experiments are performed in ultrahigh vacuum (base pressure ≤ $1.0 \times 10^{-10}$

Torr) using a commercial molecular beam epitaxy (MBE) - low temperature (LT) STM combined system (Unisoku). Bi(111) BL films were hetero-epitaxially grown on pristine $Bi_2Te_3$ films (thickness ~15 quintuple layers) supported by graphitized SiC substrates. After growth, the samples were immediately transferred into *in situ* LT-STM so that the vacancy island shape could be frozen at 4.36 K to avoid Bi migration. The mechanically sharpened PtIr tips were used for STM after calibration on standard Pb/Si(111). Point STS were performed with feedback loop open (lock-in bias modulation: 1-3 $mV_{rms}$ at 987.5 Hz), and *dI/dV* maps were obtained simultaneously with constant current images (8-10 $mV_{rms}$ at 2997 Hz). All the energies are shown with respect to the Fermi level ($E_F$). The STM images were processed using WSxM software.

**Acknowledgments**

We thank Can-Li Song, Shun-Qing Shen, Yayu Wang, Zhe-Xuan Gong, Pei-Zhe Tang for helpful discussions. This work was supported by National Science Foundation, Ministry of Science and Technology of China, and in part by the Beijing Advanced Innovation Center for Future Chip (ICFC). Y.P.J. acknowledges support from the National Thousand-Young-Talents Program. C.Z.C. acknowledges support from Alfred P. Sloan Research Fellowship and ARO Young Investigator Program Award (No. W911NF1810198).

**Author contributions**

M.C., X.C.M., and Q.K.X. conceived and designed the experiments. M.C., J.P.P, and H.M.Z carried out MBE growth and STM measurements. C.Z.C and X.F. carried out the ARPES measurements. L.L.W. and K.H. assisted the experiments. Z.G.F, F.W.Z. and P.Z did the band structure calculation for the heterostructure. Y.P.J., M.C., and X.C.M. carried out the data analyses and interpretations. Y.P.J, M.C., and X.C.M. wrote the manuscript.

**Competing interests**

The authors declare no competing interests.

**Additional information**

Correspondence and requests for materials should be addressed to Yeping Jiang, Xu-Cun Ma, Qi-Kun Xue.

**Figure Legends**

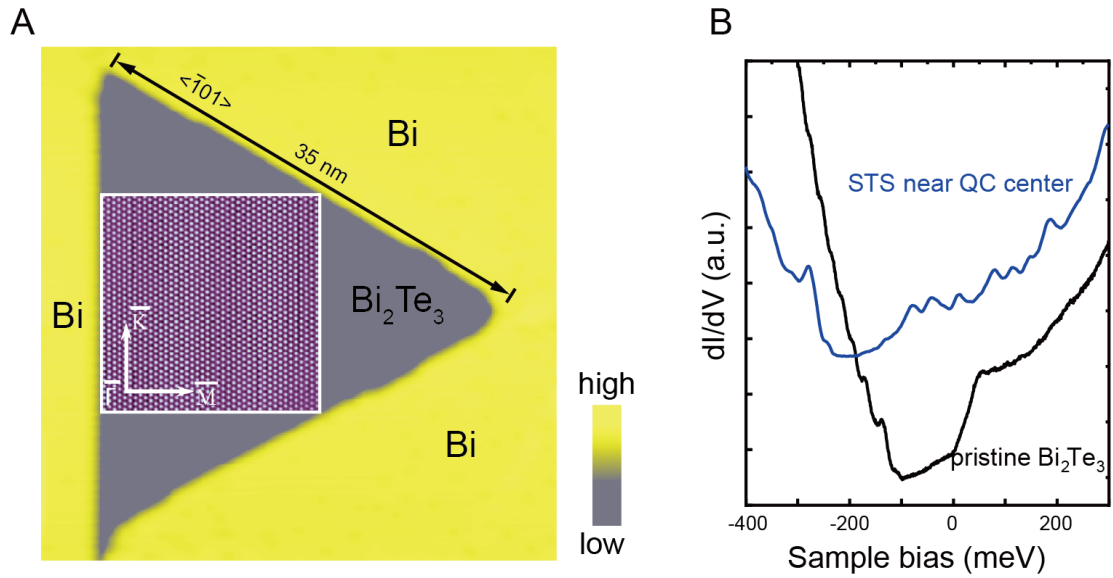

**Fig. 1. Surface topography of the 0.9 BL Bi film on Bi₂Te₃ films.** **(A)** A typical equilateral triangular Bi₂Te₃ vacancy island surrounded by Bi bilayers (brighter regions) (42×42 nm², tunneling condition: 0.5 V, 0.1 nA). The Bi bilayer was grown on the Bi₂Te₃ film by MBE. The thickness of the Bi₂Te₃ film is 15 quintuple layers. The atomic resolution image in the white square shows the defect-free Te-terminated Bi₂Te₃ (16×16 nm², -0.4 V, 0.4 nA). $\overline{\Gamma}-\overline{M}$ and $\overline{\Gamma}-\overline{K}$ in the momentum space are shown to indicate their alignment with the TQC edges. **(B)** Typical d$I$/d$V$ spectra taken near TQC center in A (upper curve) and on pristine Bi₂Te₃ (lower curve). The curves are vertically offset for clarity.

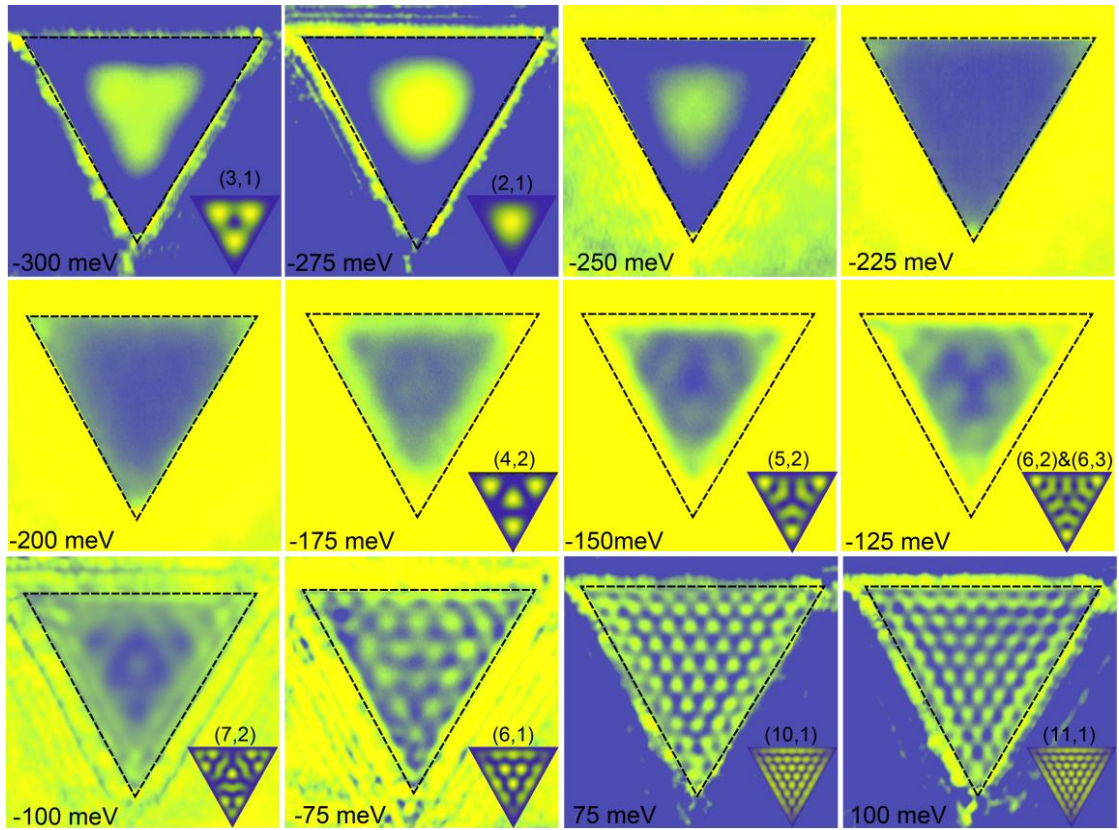

**Fig. 2. Quasi-bound states in a TQC (*L*=35 nm).** Spatial *dI/dV* maps of TQC in Fig. 1A (42×42 nm$^2$) at different energies (-300 meV, -275 meV, -250 meV, -225 meV, -200 meV, -175 meV, -150 meV, -125 meV, -100 meV, -75 meV, 75 meV and 100 meV) (see Fig. S2 for more images at other energies). The dashed triangles are guides for the eye, indicating the positon of triangular barriers. The inset in each image is the simulated pattern of certain indice which matches the interference pattern best. In this simulation, the spinor part of the wave function is omitted, which may only affect the overall intensity and does not contribute to the spatial variation of the pattern.

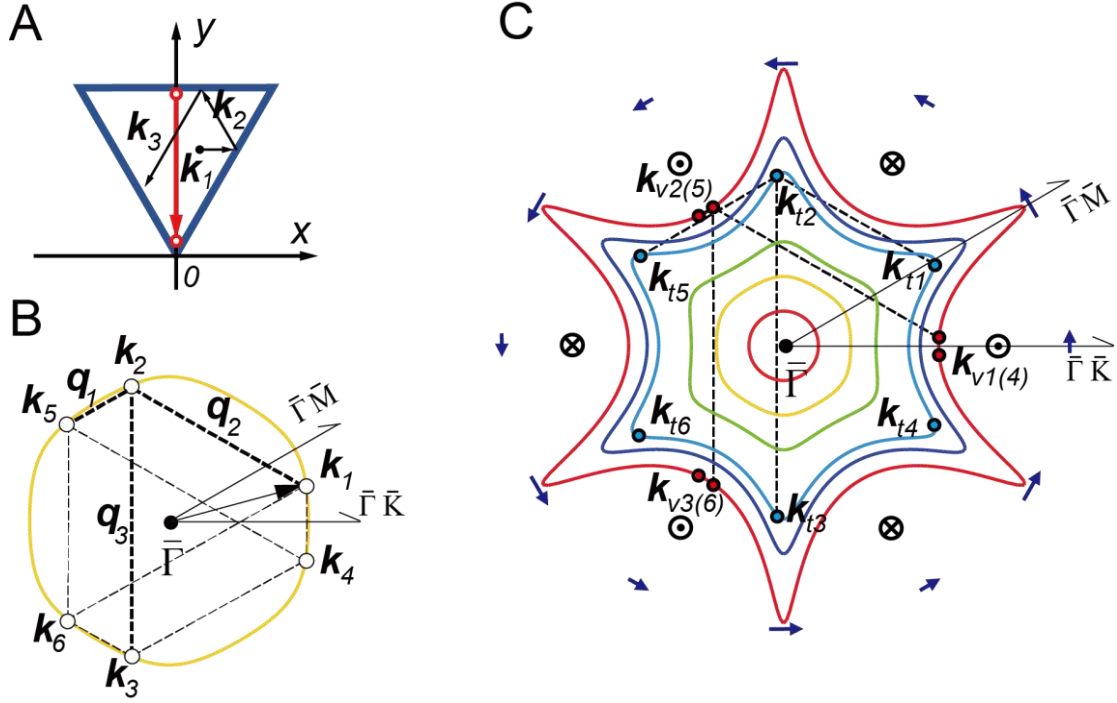

**Fig. 3. Scattering process of possible quasi-bound surface states in $Bi_2Te_3$ in a TQC. (A)** A schematic diagram of the scattering process of surface states inside the TQC. The thick (red) arrow indicates where the spatially dependent d$I$/d$V$ spectra in Fig. 4 were taken. **(B)(C)** The scattering process shown on the energy dependent CECs of TSS with a strong warping effect. The dashed lines in B indicate the possible scattering vectors between wave vectors $k_1$-$k_6$. $q_1$, $q_2$, $q_3$ denote three inequivalent reflections. $k_{ti}$ and $k_{vi}$ are the wave vectors close to the tip and valley positions on the warped CECs, respectively. At high energies, the CECs are labeled with a spin texture exclusively for TSS with a strong warping term. Note that because of the warping effect, the spin-momentum helicity deviates from those at low energies, acquiring an out-of-plane spin component in the valley positions, as indicated by the crosses and dots in C. The arrows with different lengths show the in-plane spin components with different magnitudes.

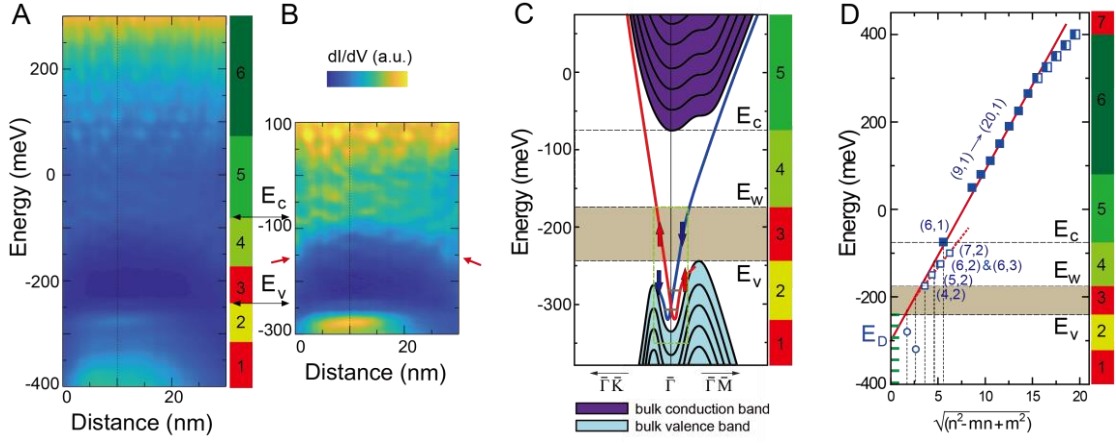

**Fig. 4. Spatially dependent energy levels inside the TQC (*L*=35 nm) and the derived spectrum of trapped TSS. (A)** Bias and spatially dependent dI/dV spectra along the line in Fig. 3A (tunneling condition: 0.2 V, 0.2 nA). The dashed line indicates the center position of the TQC. The sidebar indicates the energy regions as described in the main text having different trapping behaviors. **(B)** Blow-up maps from -300 meV to 100 meV. The red arrows indicate the localized states near the barriers. **(C)** A schematic of the band structure of $Bi_2Te_3$ thin films. The two short lines indicate the states (2, 1) and (3, 1) (see Fig. S6 for details of the green dashed region). **(D)** The spectrum ($E_{nm}$, $k'_{nm}$) of trapped TSS, where $k'_{nm} = \frac{3L}{4\pi} k_{nm}$. The half-filled squares (states ($n \geq 16$, 1)) are obtained by counting the nodes in the dI/dV maps shown in the supplementary Fig. S2. The red solid and dashed lines are the linear fitting to the states ($n \geq 6$, 1) (filled squares) and the states ($n$, $m \geq 2$) (hollow squares), respectively. The equal-spaced short lines between -400 meV and -240 meV indicate the peak positions in the dI/dV spectrum coming from the bulk valence subbands of the $Bi_2Te_3$ film. The dashed vertical lines indicate the positions of $k'_{n1}$, where $n$ ranges from 2 to 6. The gray regions in C and D denote the forbidden region for the trapping of TSS.

## SUPPLEMENTARY MATERIALS

Figure S1: Band structure of Bi/$Bi_2Te_3$ on (111) surface.

Figure S2 | Successive d$I$/d$V$ maps of the TQC (35 nm) at energies with an interval of 25 meV.

Figure S3 | d$I$/d$V$ spectra on TQCs of different sizes.

Figure S4 | d$I$/d$V$ maps on a smaller TQC (17 nm).

Figure S5 | d$I$/d$V$ spectra of Fig. 4A in the line-plot form.

Figure S6 | A zoom-in view of the schematic band structure in Fig. 4C.

Figure S7 | Section profiles of patterns at 50, 75, 100 meV along the lines indicated in Fig. S2.

Figure S8 | A particular situation adopted to give a rough estimation of the trapped states' lifetime in which the path of the electron in a TQC roughly follows a regular pattern.

# Supplementary Information

# Selective Trapping of Hexagonally Warped Topological Surface States in a Triangular Quantum Corral

Mu Chen[1,2], Yeping Jiang[3]*, Junping Peng[1], Huimin Zhang[1], Cui-Zu Chang[1,7], Xiao Feng[1], Zhenguo Fu[4], Fawei Zheng[4], Ping Zhang[4], Lili Wang[1,5], Ke He[1,5], Xu-Cun Ma[1,5]*, Qi-Kun Xue[1,5,6]*

[1] *Department of Physics, Tsinghua University, Beijing 100084, China*

[2] *Beijing Institute of Aeronautical Materials, Beijing 100095, China*

[3]*Key Laboratory of Polar Materials and Devices (MOE), Department of Optoelectronics, East China Normal University, Shanghai 200241, China*

[4] *Institute of Applied Physics and Computational Mathematics, Beijing 100088, China*

[5]*Collaborative Innovation Center of Quantum Matter, Beijing 100084, China*

[6]*Beijing Academy of Quantum Information Sciences, Beijing 100193, China*

[7]*Department of Physics, The Pennsylvania State University, University Park, Pennsylvania 16802, USA*

*To whom correspondence should be addressed to Y.P.J. (ypjiang@clpm.ecnu.edu.cn), X.C.M. (xucunma@mail.tsinghua.edu.cn) or Q.K.X. (qkxue@mail.tsinghua.edu.cn).

## Contents

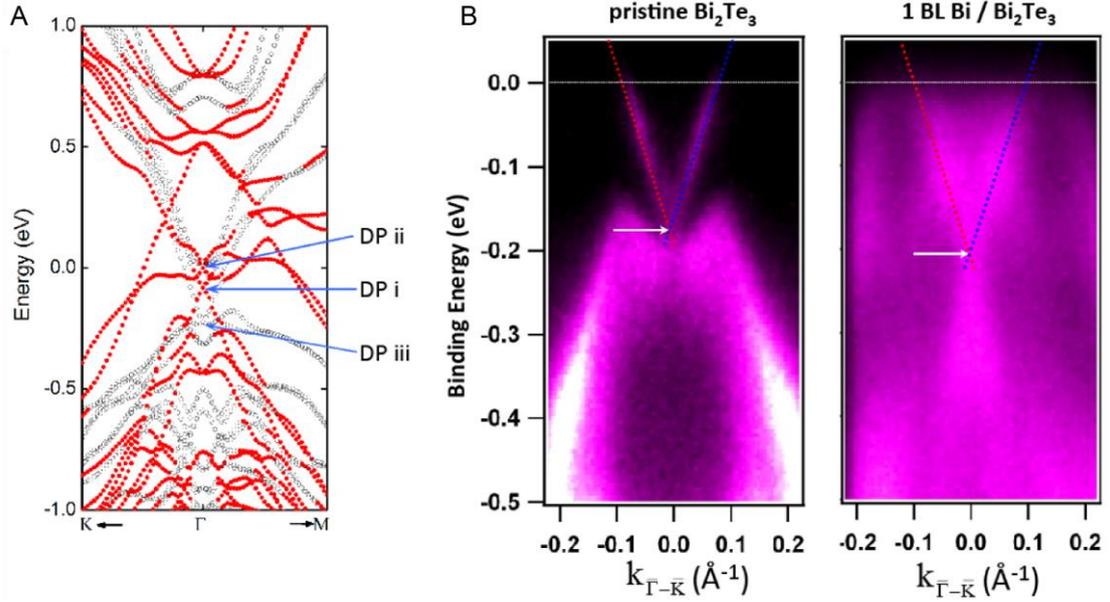

**Fig. S1. Band structure of Bi/Bi$_2$Te$_3$ on (111) surface. (A)** DFT calculations of single BL Bi on 3 quintuple layers (QLs) Bi$_2$Te$_3$. Red part depicts contributions from Bi and top QL of Bi$_2$Te$_3$. Three Dirac points coexist and are labeled. DP i from top surface of Bi$_2$Te$_3$ is reserved upon Bi deposition, survives in broad energy range and shows a sublinear dispersion along $\overline{\Gamma}-\overline{M}$ toward high energy, which is a hint of warped CEC. DP ii mainly exists on Bi films. The lowest DP iii is contributed by bottom surface of pure Bi$_2$Te$_3$ due to thickness limit. Another cone-like structure around +500 meV is actually induced by Rashba splitting on Bi. **(B)** Room temperature ARPES spectra along $\overline{\Gamma}-\overline{K}$ direction of the pristine Bi$_2$Te$_3$ films (thickness ~15 QLs), and 1 BL Bi/Bi$_2$Te$_3$ films respectively. DP i is observed and indicated by white arrows. Spin-polarized TSS are shown in red and blue dashed lines. New upward bands at large momentum around Dirac cone upon Bi deposition are contributed by Bi BL films. The surface state band structure of Bi$_2$Te$_3$ in the Bi-covered region is different from that of the uncovered region. The ARPES data (Fig. S1B) shows the band structure of the Bi-covered region. The Dirac point is no longer buried below the valence band maximum and shifts to higher energy. As we can also see in the theoretical data in Fig. S1A, the surface state band structures are now different between the top and bottom surfaces. DPi (top surface states) is lifted up by about 150 meV with respect to DPiii (bottom surface states) because of the existence of a Bi-BL overlayer. A Bi BL modifies the boundary condition for the occurrence of surface states. The combination of charge

doping and Dirac-point lifting (~ 150 meV) leads to a Dirac point at -210 meV, higher than that in the uncovered region (-300 meV).

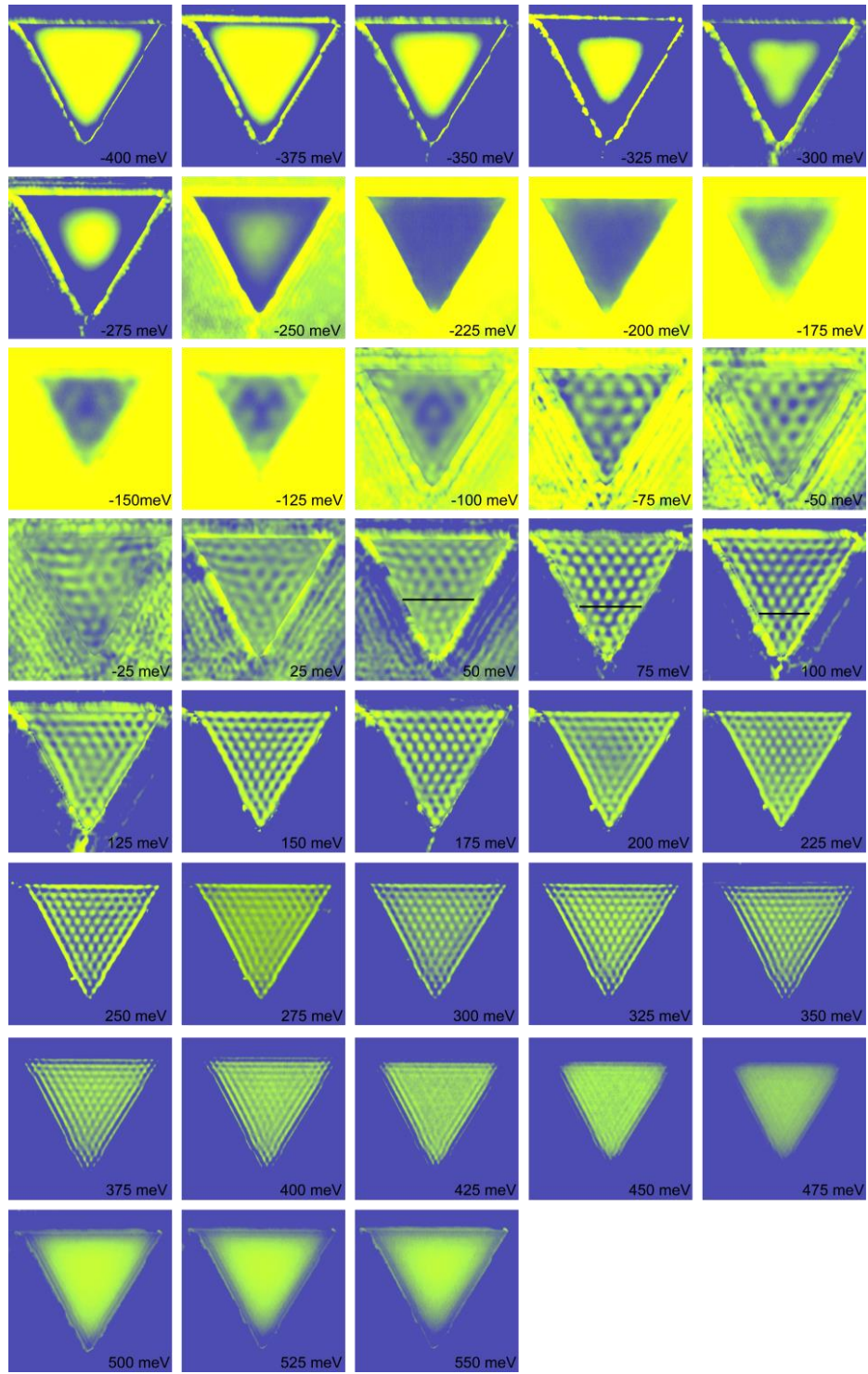

**Fig. S2. The d*I*/d*V* maps of TQC at energies from -400 meV to 550 meV with an interval of 25 meV (42×42 nm$^2$).** Spatial *dI/dV* maps of TQC in Fig. 1A of different energies. At energies between -250 meV and -175 meV no patterns can be seen. Above -175 meV there appear trapped states with different interference patterns. Above 200 meV the step-edge interference starts to dominate the pattern near the TQC edges. Above 425 meV there is no more clear patterns.

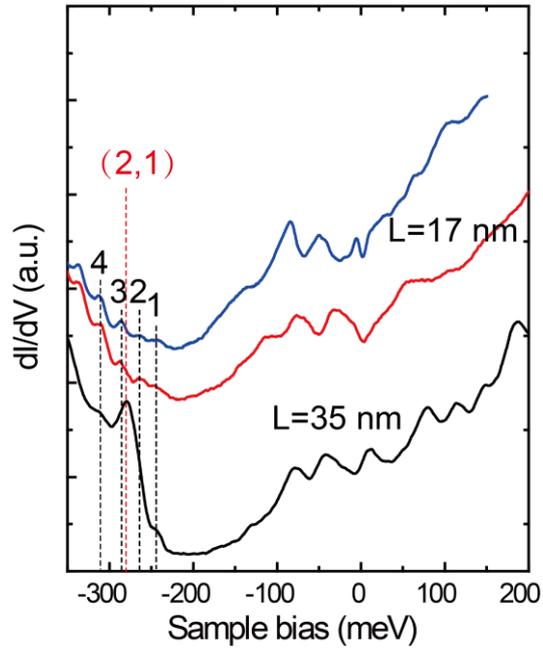

**Fig. S3. d*I*/d*V* spectra on TQCs of different sizes.** The lower curve is the d*I*/d*V* spectrum in Fig. 1D. The upper two curves are taken at different positions in a 17-nm TQC (The effective size is 25 nm as explained in Fig. S4). The curves are vertically offset for clarity.

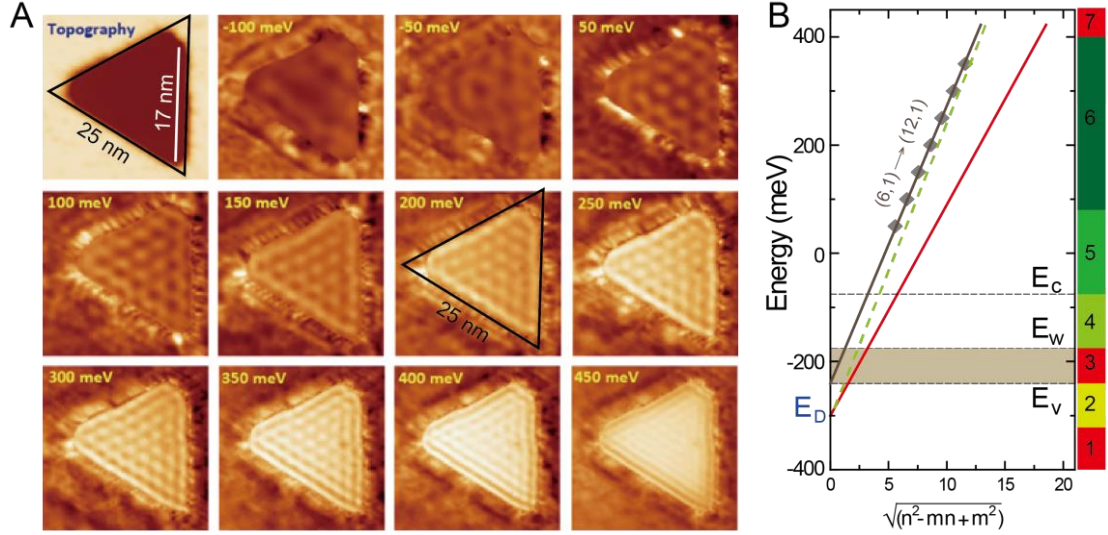

**Fig. S4. d*I*/d*V* maps (25×25 nm²) on a smaller TQC. (A)** The triangle side length *L* ~17 nm (the white line in the topographic image). Compared with the larger TQC (35 nm), the smaller one deviates more from the shape of an ideal triangle. The effective size can be derived by fitting a triangle to the interference pattern (the black triangle). This leads to an effective size of ~ 25 nm. Above 50 meV the (*n*, 1) states start to dominate the interference patterns, similar to the behaviors of the 35 nm-TQC. At 450 meV the image only has pattern from the step-edge scattering. **(B)** The spectrum of trapped states in the smaller TQC (the gray squares). The gray solid line is the linear fitting to the data. The red line is the surface state dispersion $\mathrm{E} = \hbar v_{F1} k = \frac{4\pi}{3L} \hbar v_{F1} k'$ along $\bar{\Gamma} - \bar{\mathrm{K}}$ derived from the trapped TSS in the 35-nm TQC (the red line in Fig. 4D). The green dashed line is the $\mathrm{E} - k'$ dispersion expected for the smaller TQC. Note that, as expected, the energy interval (~ 50 meV) of the trapped states for the smaller TQC is larger that that for the 35-nm one (~ 30 meV). Note that the derived Dirac energies for these two TQCs should be equal, as can be seen in Fig. S3. The discrepancy between the gray solid and green dashed lines may result from the following facts. For the smaller TQC, we donot have the data showing the spatially dependent energy levels as in Fig. 4A. Thus, the energies of the trapped states are not corrected by the spatially dependent mapping of the energy levels. Considering the trapped states may have an energy width of about 40 meV, the data without correction may lead to a huge deviation from the true dispersion.

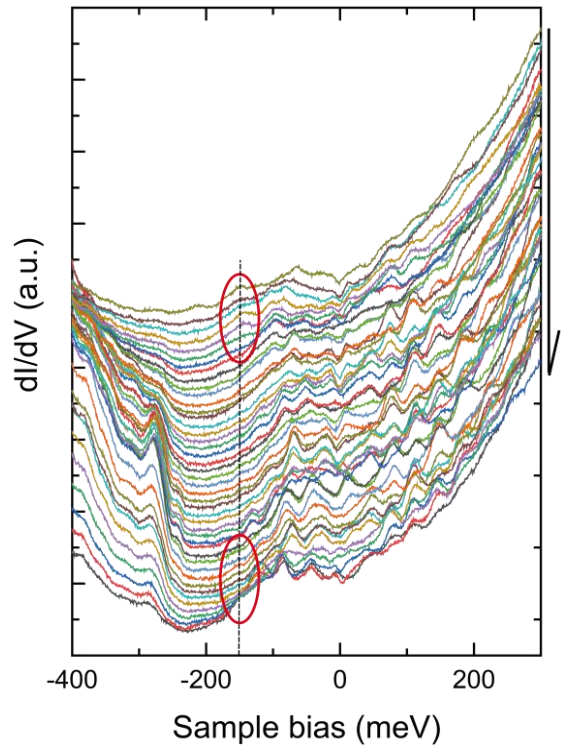

**Fig. S5. d*I*/d*V* spectra in Fig. 4A.** The arrow indicates that these spatial dependent spectra were taken along the line as shown in Fig. 3A. The two ellipses enclose the bound states close to the barrier walls.

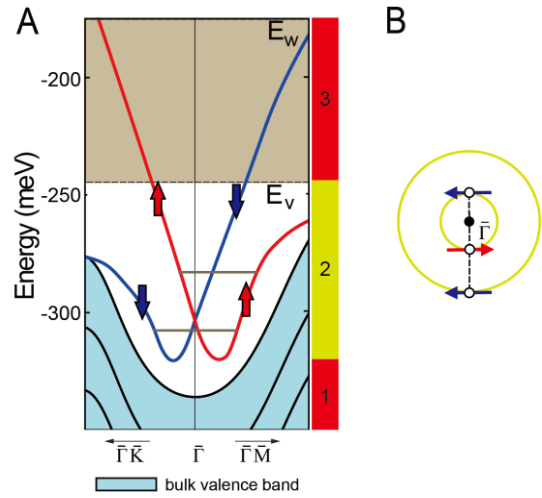

**Fig. S6. The trapping of the states (2, 1) and (3, 1). (A)** The zoom-in view of the schematic band structure (the green dashed region in Fig. 4C). The gray solid lines crossing the TSS at -280 and -320 meV indicate the states (2, 1) and (3, 1), respectively. **(B)** The CEC with double Fermi surfaces, in which backscattering is no longer forbidden.

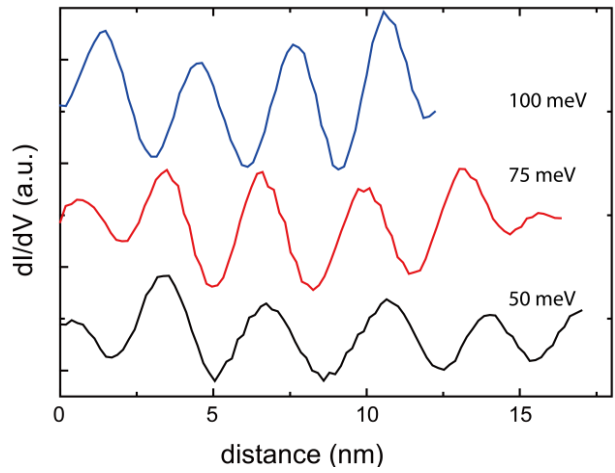

**Fig. S7. Section profiles of patterns at 50, 75, 100 meV along the lines indicated in Fig. S2.** The line-shapes near the edges indicates the different phase shift of electrons upon reflections.

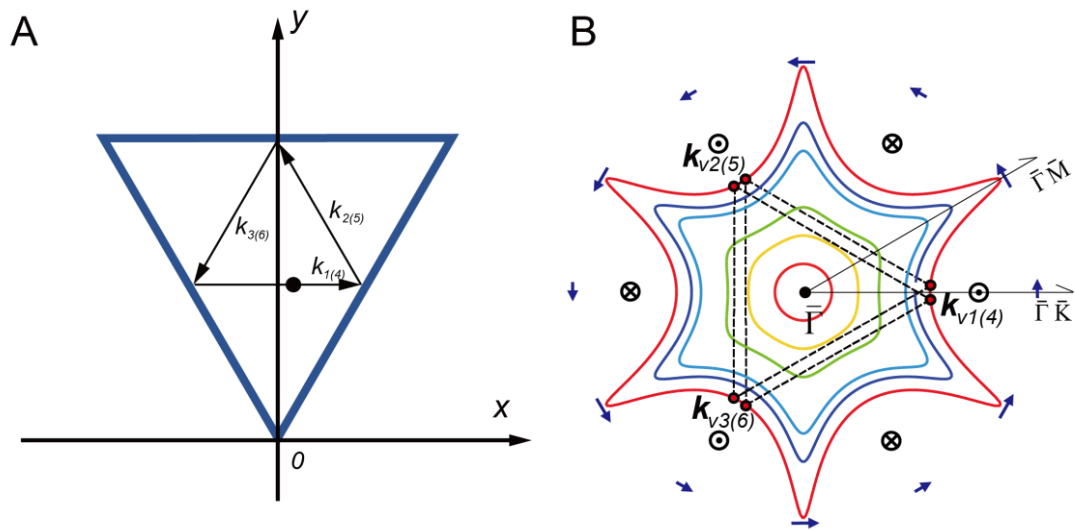

**Fig. S8. A particular situation adopted to give a rough estimation of the trapped states' lifetime, in which the path of the electron in a TQC roughly follows a regular pattern.** To get a rough estimation of the tapped states' lifetime, a specific situation is adopted. As shown in A, the valley electrons (red dots in B) close to the arrowed lines roughly follows these lines that form a triangle, at least for the first several reflections. This estimation is applicable only for the ($n$, 1) states with high $n$-indexes. The higher the index is, the closer the path is to the regular pattern. In fact, in the spectrum shown in Fig. 1B, only the peaks for the ($n$, 1) states are identified. For such a regular pattern, the time interval between reflections is fixed ($T_0=L/2/v_{F1}$). We then have the equality $R^{(\tau/T_0)} = e^{-1}$, where $R$ is the reflectivity and $\tau$ is the lifetime of the quasi-trapped states.